\documentclass[preprint,nofootinbib,tightenlines,showpacs]{revtex4}
\usepackage{amsfonts}
\usepackage{amsmath}
\usepackage{lscape}
\usepackage{epsfig}
\usepackage{graphicx}
\usepackage{enumerate}
\usepackage{xcolor}
\def\beq{\begin{equation} }
\def\eeq{\end{equation} }
\def\ba{\begin{eqnarray} }
\def\ea{\end{eqnarray} }
\def\lav{\langle\!\langle}
\def\rav{\rangle\!\rangle}
\def\nn{\nonumber }

\begin{document}

\title{Dynamical and Quenched Random Matrices and  Homolumo Gap}
\author{Ivan Andri\'c}
\email{iandric@irb.hr}
\author{Larisa Jonke}
\email{larisa@irb.hr}
\author{Danijel Jurman}
\email{djurman@irb.hr}
\affiliation {Division of Theoretical Physics, Rudjer Bo\v skovi\'c Institute,
Bijeni\v cka 54, 10000 Zagreb, Croatia}
\author{Holger Bech Nielsen}
\email{hbech@nbi.dk}
\affiliation{The Niels Bohr Institute, Copenhagen DK 2100, Denmark}

\

\vskip 2cm

\

\begin{abstract}
We consider a rather general type of matrix model, where  the matrix $M$  represents a Hamiltonian of the  interaction of a bosonic system with a single fermion. The fluctuations of the matrix
are partly given by some  fundamental randomness and partly dynamically, even quantum mechanically.
We then study the homolumo-gap effect, which means that we study how the level density for the single-fermion
Hamiltonian matrix $M$ gets 
attenuated near the Fermi surface. 
In the case of the quenched randomness (the fundamental one) dominating the quantum mechanical one
 we show that in the first 
approximation the homolumo gap  is characterized by the absence of single-fermion levels 
between two steep gap boundaries. 
The filled and empty level densities are in this first approximation just
pushed, each to its side.
In the next approximation these steep drops in the spectral density are smeared
out to have an error-function shape.
The studied model could be considered as a first step towards the more
general case of  considering a whole field of matrices - defined say on some phase space - 
rather than a single matrix.
\end{abstract}

\maketitle

\section{Introduction}
The homolumo gap goes back to the Jahn-Teller effect \cite{jt}, which describes
spontaneous  breaking of the molecular symmetry owing to the partial filling of
degenerate electron states in the molecular ``shell'' at the Fermi energy.
In general, it is not difficult to see that if some deformation of the molecule
could lower the filled and raise the empty levels this would lower the energy of
the electrons \cite{jt2}. Thus an attenuation of the density of electron levels near the Fermi
level is expected, i.e. the interval from the \underline{h}ighest \underline{o}ccupied
level (\underline{m}olecular \underline{o}rbit) ''homo'' up to the \underline{l}owest 
\underline{u}noccupied level (\underline{m}olecular \underline{o}rbit), ''lumo'',  
would be increased. It is from these short hand notations ''homo'' and ''lumo'' that 
the effect has got the name homolumo effect and the spacing between the two
levels neighboring the Fermi level the name homolumo gap.  
The point giving an interest to the homolumo effect is that it is so general
that one might find in nature many systems showing such effect.

In the field of material sciences the  study of the
material-specific characteristics of the homolumo gap attracts a lot of
attention, see Ref.\cite{Kim} as an example.
However, dependence of the homolumo gap on the relevant parameters in a specific
system may be very complicated due to complexity of interactions involved.
Therefore, to study formation of the homolumo gap in a set up not so   dependent
on the details of the interactions one would  make some simplified model which
captures the main features of such complex systems. 
An obvious simplified system to think about is a molecule in which positions of
the nuclei and the collective charge distribution of the electrons produce 
a Hamiltonian for a single electron. 
We assume that these complicated molecule interactions can be described 
by a matrix. Accordingly, we consider the  matrix elements as dynamical variables due to
their dependence on the positions of the nuclei and other degrees of freedom.
Furthermore, we assume that there exists some suitable stable ground state of the molecule, 
such that the dynamics of the matrix elements can effectively be described by a
harmonic approximation.

In the previous  works \cite{hl1,hl2} we have introduced and analyzed such a model. Schematically, the model is defined by matrix Hamiltonian $H=H_B+gH_{FB}$, where $H_B$ is Hamiltonian of bosonic subsystem, while $H_{FB}$  represent the interaction Hamiltonian with  constant $g$ controlling the coupling strength. 
{We have been able to find the approximate ground state  giving rise to
the homolumo gap in the limit of strong interaction  between fermions and bosonic matrix.
This limit is characterized by the size of the homolumo gap being much larger then the scale of spread of the distribution of the 
filled/empty levels. 
In particular, the approximate ground state  in the strong coupling limit was obtained by discarding the part of Hamiltonian which 
describes transition between filled and empty levels and 
scales as $1/g^2$ \cite{hl2}.}

 {In the present work we modify the model by subjecting the parameters determining the ground state to some distribution law.
Thus we introduce one more parameter in the model which allows us to effectively freeze the dynamics of bosonic degrees of freedom and to 
obtain homolumo gap even in the case when displacement of levels is comparable with the scale of the distribution.}  
From the physical perspective, subjecting the parameters of the ground state to a distribution law  could be justified in several ways. 
For example one could simply assume that there is some  fundamental randomness  in
the system or one could in that way   describe some external influence.
Moreover, the system which we consider might be so complicated that we  do not want to 
or cannot treat it in detail. We rather make   statistical ansatz for the hard to compute 
properties of the system in order to obtain   trustworthy treatment of certain general features 
of the system. Since  the Hamiltonian for a single fermion can be represented as a
matrix it is natural to consider this matrix having a random distribution, 
but we must nevertheless consider it dynamical, so that it can be acted upon and 
pushed by forces from the fermions.  

In this paper we build the model on these general assumptions. In its ground
state the system can be viewed as a particular representative of the models  
known in the literature as the ``random matrix models in an external field''
\cite{Zinn-Justin, Bleher}.
We believe that by considering various forms of randomness and dynamics in the 
model one should be able to approach  close to specific realistic situations 
and describe some general features, like the formation of the homolumo gap,
reasonably correctly. 

In the following  section we shall set up our model at first w.r.t. the bosonic
degrees of freedom described by the dynamical matrix $M$ and show how  
inclusion  of the random matrix $M_0$ affects the ground state. 
Then in section \ref{adding} we introduce $N_f$  fermions interacting with the matrix $M$, 
so that the matrix itself becomes the single-fermion Hamiltonian. For a certain range of parameters of the model we find the
ground-state wave function. 
In  section \ref{largeg} the density of the lowest $N_f$ single-fermion levels in the ground state 
is studied and the effect of
back-reaction of the fermions on the matrix dynamics is considered, so that a true homolumo gap can be derived. 
In section \ref{conclusion} we discuss the importance of the obtained results 
and some plans for further work.

\section{The density of the lowest $N_f$ levels in the free matrix model}
\label{sec2}

As explained in introduction, we want to set up a model describing dynamical system 
of interacting bosons and fermions, where some of the model
complexity is contained in the random distribution of model parameters. 
In the bosonic subsystem this means that we want to set up a dynamical $N\times N$
matrix $M$ whose matrix elements fluctuate according to two different laws. 
First,  we assign the probability distribution to matrix elements of 
a fundamental $N\times N$ random matrix $M_0$ giving what we call the central values of the dynamical matrix $M$
and choose  general symmetry properties (e.g. under transposition, such as the 
matrix being hermitian).
Thus we start with a hermitian random matrix $M_0$ whose  distribution 
is just of the type studied as the simplest case in \cite{mehta}:
\begin{equation}
{\cal P}(M_0)={\cal N}(\tilde
\omega,N)\exp{\left(-{\tilde\omega}\mathrm{Tr}M_0^2\right)},
\label{qdistribution}
\end{equation}
where the normalization constant is given by:  
\beq
{\cal{N}}(\tilde\omega,N)=2^{N(N-1)/2}
\left(\frac{\tilde\omega}{\pi}\right)^{N^2/2}.
\eeq
It is well known that its eigenvalues have the famous semicircle
distribution in the large-N limit, see (\ref{semicircle}) below.

Secondly, we want the matrix to be dynamical in the sense that it can be
influenced by interaction with fermions. These dynamical degrees of freedom 
can then be quantized, resulting in the fluctuations in the ground state.
In our picture the matrix $M_0$  contains all complicated information about the system such as for example  the value of the minimum of the
potential energy in the system.
Assuming that the system has such minimum, we expect that low energy spectrum 
is well described by harmonic approximation. Accordingly, we use the random matrix $M_0$ to define potential energy
for a dynamical  matrix $M$:
\begin{equation}
V(M) = \frac{\omega^2}{2} \mathrm{Tr}(M-M_0)^2.
\label{potential}
\end{equation} 
Then the simplest is to supplement this potential energy with the kinetic energy
that is usual in matrix models to obtain the Hamiltonian for our matrix model
\begin{equation}
H_{HO}=\frac{1}{2}\mathrm{Tr} P^2+\frac{\omega^2}{2}\mathrm{Tr}(M-M_0)^2,\; 
\left[P_{ij},M_{kl}\right]=-i\delta_{ik}\delta_{jl}.
\label{hamM}
\end{equation}
Although in this paper we work with hermitian matrices,
generalization to e.g. quaternionic hermitian ones is straightforward. 

The ground state wave-function of the system defined by the Hamiltonian 
(\ref{hamM}) is
\begin{equation}
\Psi_0^{gs}(M)=\sqrt{{\cal N}(\omega,N)}\exp{\left(-\frac{\omega}{2}\mathrm{Tr}(M-M_0)^2\right)}.
\label{psiM}
\end{equation}
The expectation value of the level density $\rho(x)=\mathrm{Tr}\delta(x-M)$ in
the ground state (\ref{psiM}) averaged over the ensemble of $M_0$ is 
\beq
\lav\rho(x)\rav_{\mathrm{0}}^{gs}=
{\cal{N}}\int\int dM dM_0 e^{-{\tilde\omega}\mathrm{Tr}M_0^2}
e^{-{\omega}\mathrm{Tr}(M-M_0)^2}\mathrm{Tr}\delta(x-M)=e^{-\omega_r x^2}\sum_{i=0}^{N-1}
\frac{H^2_i\left(\sqrt{\omega_r}
x\right)}{ 2^i i!\sqrt{\pi}}, 
\label{gsdens}
\eeq
where we choose normalization 
${\cal N}={\cal N}(\omega,N){\cal N}(\tilde\omega,N)$
so that the level density is normalized as
\beq\label{NN}
\int dx \rho(x)= N,
\eeq
where $N$ is the number of levels, i.e. the order of the matrices  $M$ and
$M_0$. Parameter  $\omega_r$ defined by 
\begin{equation}
\omega_r=\frac{\omega\tilde\omega}{\omega+\tilde\omega},
\end{equation} 
characterizes spread of the expectation value of the level density.
In the large-$N$ limit, the weighted sum of squared Hermite polynomials which appears in (\ref{gsdens})
reduces to the Wigner semicircle distribution \cite{mehta}:
\begin{eqnarray}\label{semicircle}
\lav\rho(x)\rav_{\mathrm{0}}^{gs}&\approx&
\frac{\omega_r}{\pi}\sqrt{ \frac{2N}{\omega_r}
 - x^2}, \end{eqnarray}
non-vanishing only for $x\in[-\sqrt{2N/\omega_r},\sqrt{2N/\omega_r}]$.
Localization of the eigenvalues described by Wigner semicircle law can 
be seen as a result of the competition between attractive harmonic oscillator potential and the 
repulsion between levels effectively induced by Vandermonde determinant 
which appears in the integration measure of $dM$ and prevents degeneracy, i.e., the  eigenvalues crossing.
We notice that as a consequence of introducing the probability distribution (\ref{qdistribution}) 
for the parameters of the model (\ref{hamM}) the width of the distribution grows as $\omega_r\leq\omega$.

Up to now we used ``double averaging'' 
notation $\lav O \rav_0^{gs}$ to stress that the averaging should be performed by taking  the expectation 
value in the ground state with the fixed matrix $M_0$, but also to average over the G(aussian)U(nitary)E(nsemble) of 
$M_0$, and subscript $0$ indicates that we put fermion-boson interaction to zero.  
In the rest of this  section we denote these averages by a single symbol $\langle O \rangle$ assuming that 
average over the GUE of $M_0$ is performed.

In the following we are interested in the expectation value of density of the $N_f$
lowest eigenvalues of matrix $M$ defined by:
\ba\label{dennf}
\rho_{N_f}^<(x)=\sum_{\cal{C}}\prod_{i,\alpha}
\theta(\lambda_{{\cal{C}}(\alpha)}-\lambda_{{\cal{C}}(i)})\sum_{{\cal{C}}(i)}
\delta(x-\lambda_{{\cal{C}}(i)}),
\ea
where the sum over ${\cal C}$ denotes the  sum over all possible 
combinations of dividing $N$ eigenvalues into the $N_f$ lowest 
(denoted by indices ${\cal{C}}(i)=1,\cdots,N_f$) and  $N_e=N-N_f$ highest 
(denoted by indices ${\cal{C}}(\alpha)=N_f+1,\cdots,N$). 
Using  that integration measure over the matrix configuration space vanishes on the 
subspace of degenerate eigenvalues the  density (\ref{dennf}) can be rewitten as 
\begin{equation}
\rho_{N_f}^<(x)=\theta (\lambda_{N_f}-x)\rho(x).
\end{equation} 
In order to calculate the expectation value of this density we need the distribution $\langle \delta(x-\lambda_{k}) \rangle$
 of  a particular eigenvalue $\lambda_k$ chosen from the ordered set of eigenvalues.
 We recall that this  distribution is localized
around some constant value, which can be approximated by the $k$'th zero of the  Hermite polynomial $H_N (x)$ 
which minimizes the  potential $W$ \cite{mehta,gustavson}:
\ba
W=\frac{1}{2} \sum_{i} \lambda_i^2-\sum_{i<j} \ln{|\lambda_i-\lambda_j|}.
\ea
This potential is  (logarithm of)   the product of the probability (\ref{qdistribution}) and Vandermonde determinant arising from the
integration measure and thus  it's minimum determines the most probable configuration.
In the large $N$ limit  in the bulk, the $k$'th zero of the  Hermite polynomial $H_N(x)$ can be replaced by $\mu_k$ \cite{DKMV} defined by:
\ba\label{flev}
k=\int_{-\infty}^{\mu_k} dx \langle \rho(x) \rangle.
\ea
Moreover, in Ref.\cite{gustavson} it has been shown that  distribution $\langle \delta(x-\lambda_{k}) \rangle$ 
is of the Gaussian form and we can write:
\ba\label{treba}
\langle \delta(x-\lambda_{k})\rangle= \sqrt{\frac{\omega_k}{\pi}}
e^{-\omega_k(x-\mu_k)^2}.
\ea
 The parameter $\omega_k$ is determined \cite{mehta,gustavson}  by fluctuations of the 
number of eigenvalues in the interval  $<-\infty,\mu_k]$ and is given by
\ba\label{om}
\omega_k\approx\frac{\pi^2\langle \rho(\mu_k)\rangle^2}{\ln N}.
\ea
Now we are in a position to find the expectation value of density (\ref{dennf}).

\begin{figure}[h]
\includegraphics[scale=1,width=8cm]{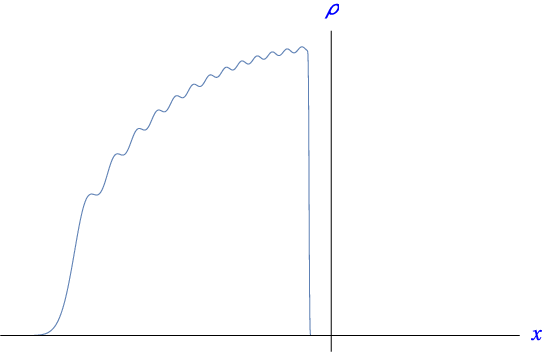}
\includegraphics[scale=1,width=8cm]{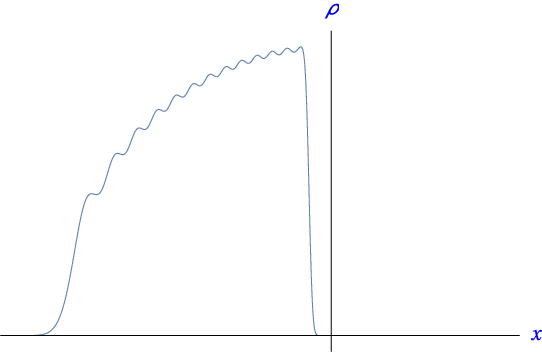}
\caption{We sketched the $N_f$ lowest levels density $\rho=\lav \rho_{N_f}^<(x)\rav_{\mathrm{0}}^{gs}$ 
for $\omega_F\to \infty$ (left)
and for $\omega_F=9.1$, taking $N=28$ and $\mu_F=-0.62$ (giving $N_f=13$).}
\label{fig2}
\end{figure}

We would like to use approximation:
\begin{equation}\label{approximation}
\langle \theta(x-\lambda_{N_f}) \rho(x)\rangle \approx \langle \theta(x-\lambda_{N_f})\rangle \langle \rho(x)\rangle.
\end{equation}
To justify this approximation we introduce the smeared density of levels  obtained 
by  replacing the delta functions in the definition of $\rho(x)=\sum_i  \delta(x-\lambda_i)$ by
the finite-width Gaussian distributions   
\begin{equation}\label{Om}
\rho_\Omega (x)=\sqrt{\frac{\Omega}{\pi}} \sum_{i=1}^N e^{-\Omega(x-\lambda_i)^2}
=\sqrt{\frac{\Omega}{\pi}}\int_{-\infty}^{\infty} dy  e^{-\Omega(x-y)^2} \rho(y),
\end{equation}
where we are free to choose $\Omega$ depending on the required accuracy.
Here we have in mind that replacing the densities with smeared ones actually means integration of 
original correlation function over intervals of the width of smearing $\Delta x\sim  1/\sqrt{\Omega}$. 
Thus we take  the density-density correlation function obtained in Ref.\cite{bz}:
\ba\label{corzee}
&&\langle \rho(x)\rho(y)\rangle-\langle \rho(x)\rangle \langle\rho(y)\rangle=
\langle \rho(w)\rangle^2 \left(\pi \delta(z)-\frac{\sin^2 z}{z^2}\right),\;w=\frac{x+y}{2},\; 
z=\pi \langle \rho(w) \rangle (x-y),\nn
\ea
and  replace $\rho(x)$ by $\rho_\Omega (x)$ to obtain
\ba
&&\langle \rho_\Omega(x)\rho_\Omega(y)\rangle-\langle \rho_\Omega(x)\rangle \langle\rho_\Omega (y)\rangle=\\
&&=-\frac{\Omega}{\pi}
\int_{-\infty}^{\infty} dw^\prime e^{-2{\Omega}(w-w^\prime)^2}
\langle \rho(w^\prime)\rangle 
 \int_{-\infty}^{\infty} dz^\prime e^{-\tilde{\Omega} (z-z^\prime)^2} 
\left(\frac{\sin^2 z^\prime}{\pi z^{\prime 2}}-
\delta(z^\prime)\right),\;
\tilde{\Omega}=\frac{\Omega}{2\pi^2\langle\rho(w)\rangle^2}.\nn
\ea
Using the approximation
\ba 
&&\int_{-\infty}^{\infty} dz e^{-\tilde{\Omega} (z^\prime-z)^2} 
\frac{\sin^2 z}{\pi z^2}\approx
\int_{z^\prime-\frac{1}{\sqrt{\tilde{\Omega}}}}^{z^\prime+
\frac{1}{\sqrt{\tilde{\Omega}}}}dz \frac{\sin^2 z}{\pi z^2}\leq \int_{-\frac{1}{\sqrt{\tilde{\Omega}}}}^{\frac{1}{\sqrt{\tilde{\Omega}}}}dz \frac{\sin^2 z}{\pi z^2}\approx
\int_{-\infty}^{\infty}dz \frac{\sin^2 z}{\pi z^2}-\nn\\
&&-
\int_{\frac{1}{\sqrt{\tilde{\Omega}}}}^{\infty}dz
\frac{1}{\pi z^2}=1-\frac{\sqrt{\Omega}}{\sqrt{2}\pi\langle\rho(w)\rangle},\nn
\ea
we find: 
\ba\label{correlation}
\left|\langle \rho_\Omega(x)\rho_\Omega(y)\rangle-\langle \rho_\Omega(x)\rangle \langle\rho_\Omega (y)\rangle\right|\leq
\frac{\Omega^{\frac{3}{2}}}{\sqrt{2}\pi^2}
\int dw e^{-2{\Omega}(w^\prime-w)^2}=
\frac{\Omega}{2\pi\sqrt{\pi}}.  
\ea
Applying the Cauchy-Schwarz-Bunyakovsky inequality
\ba 
\left|\langle f(x)g(x) \rangle - \langle f(x)\rangle \langle g(x) \rangle\right|^2 \leq
\left(\langle f^2(x) \rangle - \langle f(x) \rangle^2 \right)
 \left(\langle g^2(x) \rangle - \langle g(x) \rangle^2 \right),\nn
\ea
on distributions $f(x)=\theta(x-\lambda_{N_f})$ and $g(x)=\rho(x)$ 
and using (\ref{correlation}) we obtain:
\ba\label{scbi} 
\left| \langle \theta(x-\lambda_{N_f}) \rho(x) \rangle - 
\langle \theta(x-\lambda_{N_f}) \rangle \langle \rho(x) \rangle \right|^2\leq \frac{\Omega^2}{4\pi^3}.
\ea 
Recall here that our analysis is performed in the large-$N$ limit and for the bulk eigenvalues, i.e., far from the edges of distributions.
Therefore, taking $\omega_r\sim N$ and choosing $\ln \Omega/\ln N < 2$ we see that we can use approximation (\ref{approximation}):
\ba 
\langle \theta(x-\lambda_{N_f}) \rho(x) \rangle \approx  
\langle \theta(x-\lambda_{N_f}) \rangle \langle \rho(x) \rangle
=\int dy \theta(y-x)\langle \delta(y-\lambda_{N_f}) \rangle \langle \rho(x) \rangle.
\ea
Therefore, the expectation value of density of $N_f$ lowest eigenvalues 
in the ground state (using again double averaging notation) is given by 
\ba 
\lav \rho_{N_f}^<(x)\rav_{\mathrm{0}}^{gs}
&\approx&\sqrt{\frac{\omega_F}{\pi}}
\int d\mu e^{-\omega_F
(\mu-\mu_F)^2}\theta(\mu-x)\lav\rho(x)\rav_0^{gs}=\nn\\
&=&\frac{1}{2}{\rm erfc}(\sqrt{\omega_F}
(x-\mu_F)) \lav\rho(x)
)\rav_{0}^{gs}~,\label{final}
\ea
where $\mu_F\equiv \mu_{N_f}\approx \langle\lambda_{N_f}\rangle$.
In the last expression (\ref{final}) we obtained the product of the density
distribution (\ref{semicircle}) and the complementary error function 
$\textrm{erfc}(z)=1-\textrm{erf}(z)$, for $z=\sqrt{\omega_F}(x-\mu_F)$.
Notice that the effect of smoothing-out of the edges of the
distribution characterized by $\omega_F$ (\ref{om}) is not the consequence of  
  the smearing introduced by $\Omega$ (\ref{Om}) since we  chose $N^2> \Omega \gg N^2/\ln N$.
In figure (\ref{fig2}) we sketch the level density of the lowest $N_f$ 
levels for fixed and fluctuating Fermi level.
\noindent Density of the highest $N-N_f$ levels is similarly obtained as 
\ba 
\lav\rho_{N-N_f}^>(x)\rav_0^{gs} &=&\frac{1}{2}\left(1+{\rm
erf}(\sqrt{\omega_F}
(x-\mu_F))\right) \lav\rho(x)\rav_0^{gs}.
\ea

\section{The lowest-energy state in the model with bosons coupled to fermions}
\label{adding}

In this section we add fermions to the system  which 
interact with the dynamical matrix $M$ functioning as the single-fermion Hamiltonian.
Similar model has already been studied in our previous papers \cite{hl1,hl2}. There
we showed that the lowest-energy state of the system with $N_f$
fermions  described by the Hamiltonian:
 \begin{equation}
 H=\frac{1}{2} \mathrm{Tr} P_M^2+\frac{\omega^2}{2} \mathrm{Tr} M^2+\frac{g}{2}
\sum_{ij} M_{ij} 
 (f^\dagger_i f_j- f_j f^\dagger_i),
 \end{equation}
in the strong coupling limit can be approximated  by  
\begin{equation}
|\psi_{gs}^{N_f} \rangle\sim e^{-\frac{\omega}{2} \mathrm{Tr}
\left(M+\frac{g}{2\omega^2}V \eta V^\dagger
 \right)^2} \tilde{f}_{i_1}^\dagger\cdots
\tilde{f}_{i_{N_f}}^\dagger|0\rangle_f,\;\eta=\left[
  \begin{array}{cc}
 \mathbb{I}_{N_f}  & 0\\
  0 & -\mathbb{I}_{N-N_f}\\ 
  \end{array}\right].
\end{equation}
In the above expression $V$ is unitary matrix which diagonalizes the matrix $M$, 
the fermions are written in the new basis  $\tilde{f}=V^\dagger f$ and
$\mathbb{I}_{N_f}$ and $\mathbb{I}_{N-N_f}$ are unit block matrices of size $N_f\times N_f$ and $(N-N_f)\times (N-N_f)$ respectively.

In this paper we modify the described model  by introducing the 
random matrix $M_0$ which specifies the  potential for the matrix $M$ 
and we define system by the following Hamiltonian:
\ba\label{HH}
&& H_{Q}=\frac{1}{2} \mathrm{Tr} P_M^2+\frac{\omega^2}{2} \mathrm{Tr}
(M-M_0)^2+\frac{g}{2} \sum_{ij} M_{ij} 
 (f^\dagger_i f_j- f_j f^\dagger_i).
\ea
Our expectation is that by appropriate tuning of the model parameters we could gain access 
to the previously \cite{hl1,hl2} inaccessible limit  in which the range of displacement $g/\omega^2$ 
 is comparable with the spread of the eigenvalue distribution.

We consider the lowest energy state of Hamiltonian $H_Q$, with fixed number $N_f$ of fermions, where $N_f$ is 
eigenvalue of the fermion number operator $\hat{N}_f=\sum_i f^\dagger_i f_i$ which is conserved quantity.
The basis of the space of states on which Hamiltonian $H_{Q}$ acts can be built 
by acting with boson and fermion creation and annihilation operators
on the ground state with $N_f=0$:
\begin{equation}
   |\psi_{gs}^0\rangle \sim e^{-\frac{\omega}{2} \mathrm{Tr}
(M-M_0-\frac{g}{2\omega^2})^2} |0 \rangle_f.
\end{equation}  
It follows that all states contain the Gaussian factor $\exp(-\frac{\omega}{2} \mathrm{Tr} (M-M_0)^2)$.
Furthermore, assuming  that $\tilde{\omega}\ll\omega$ i.e. that the spread
of random distribution is much larger than the spreading due to quantum fluctuations and 
at the same time keeping large quantity  $g/\omega\sqrt{\tilde{\omega}}\gg1$ we are able to 
estimate contributions to the $H_Q$. For that purpose we write $H_{Q}$ as a sum of three 
contributions:
\ba\label{modHH}
&&H_Q=H_{HO}+H_{F}+H_{FB},\\
&& H_{F}=\frac{g}{2}\sum_{ij} {M_0}_{ij}(f^\dagger_i f_j- f_j
f^\dagger_i),\;H_{FB}=\frac{g}{2}\sum_{ij} (M-M_0)_{ij}(f^\dagger_i f_j- f_j f^\dagger_i).
\ea
Expected order of contributions to the energy from the quantum oscillations $H_{HO}$ 
and the fermion-boson interaction $H_{FB}$ are $\omega$ and $g/\sqrt{\omega}$ respectively 
and dominant term in $H_Q$ is $H_{F}$, which is of order $g/\sqrt{\tilde{\omega}}$.
Taking this into account we define transformed fermion 
operators: $f\to \tilde{f}=V^\dagger_c f$,  where $V_c$ is unitary matrix which diagonalizes matrix $M_0\to V^\dagger_c M_0 V_c=D$.
In this basis the Hamiltonian $H_F$ takes simple form 
$H_F=\sum_i d_i(\tilde{f}^\dagger_i \tilde{f}_i- \tilde{f}_i \tilde{f}^\dagger_i)$.

We also transform  matrix $M$, $M\to V^\dagger_c M V_c=\tilde{M}$ and we split Hamiltonian $H_Q$ into two parts:
\ba
&& H_Q=H_0+\Delta H,\\
&&H_{0}=H_F+
\frac{1}{2} \mathrm{Tr} P_{\tilde{M}}^2
+\frac{\omega^2}{2} \mathrm{Tr}
(\tilde{M}-D+\frac{g}{2\omega^2}F)^2
-\frac{g^2}{8\omega^2}N,\;
 F_{ij}=\delta_{ij} 
 (\tilde{f}^\dagger_i \tilde{f}_j- \tilde{f}_j \tilde{f}^\dagger_i),\\
&&
\Delta H=g
\sum_{i\neq j} \tilde{M}_{ij} \tilde{f}^\dagger_i \tilde{f}_j.
\ea 
The Hamiltonian $H_0$ is unitarily equivalent to the Hamiltonian $H_F+H_{HO}$ which describes system of decoupled fermions and bosons:
\ba
H_0=U(H_F+H_{HO})U^\dagger,\;U=
e^{-i\frac{g}{\omega2}\mathrm{Tr} P_{\tilde{M}}F}.
\ea
The eigenstates of the Hamiltonian $H_F+H_{HO}$ are given in terms 
of usual creation and annihilation operators 
$\tilde{B}^\dagger_{ij}=(1/\sqrt{2\omega}){P_{\tilde{M}}}_{ij}+i(\sqrt{\omega}/{2})(\tilde{M}-D)_{ij}$ and $\tilde{f}^\dagger_i$ 
acting on the vacuum state $|0\rangle \sim \exp(-\frac{\omega}{2}\mathrm{Tr}(\tilde{M}-D)^2)|0\rangle_f$ 
and consequently, the eigenstates of the Hamiltonian $H_0$ can be built by acting with 
the transformed creation and annihilation operators 
\begin{equation}
U\tilde{B}^\dagger_{ij}U^\dagger=
\tilde{B}^\dagger_{ij}-i\frac{g}{(2\omega)^{\frac{3}{2}}}F_{ij},
\;U\tilde{f}^\dagger_iU^\dagger=e^{-i\frac{g}{\omega^2}P_{ii}}\tilde{f}^\dagger_i,
\end{equation}
on the transformed vacuum state.
Then the non-degenerate, lowest-energy state of the Hamiltonian $H_0$
with fixed number of fermions is given by 
\begin{equation}\label{gs}
|\psi_{gs}^{N_f}(0)\rangle\sim e^{-\frac{\omega}{2}
\mathrm{Tr} (\tilde{M}-D+\frac{g}{2\omega^2} F)^2}\tilde{f}_1^\dagger
\tilde{f}_2^\dagger 
\cdots \tilde{f}_{N_f}^\dagger|0\rangle_f,
\end{equation} 
and has energy:
\ba 
E_{gs}^{(0)}=\frac{\omega N^2}{2}-\frac{g}{2}\sum_{i=1}^{N_f}d_i+\frac{g}{2}
\sum_{i=N_f+1}^N d_i,
\ea
provided that the unitary matrix $V_c$ is chosen in such way that the
eigenvalues of matrix $M_0$ are ordered $d_i<d_\alpha, i=1,\cdots,N_f,\;\alpha=N_f+1,\cdots,N$.
Other eigenstates of $H_0$ with fixed number of fermions 
are given by applying the operators
\begin{equation}
\tilde{B}^\dagger_{ij}-i\frac{g}{(2\omega)^{\frac{3}{2}}}F_{ij},\;
e^{i\frac{g}{\omega^2}(P_{ii}-P_{jj})}\tilde{f}^\dagger_i \tilde{f}_j,\nn
\end{equation}
on the ground state $|\psi_{gs}^{N_f}(0)\rangle $.

The accuracy of the approximation in which we replace 
the lowest-energy state of the Hamiltonian $H_Q$ 
with fermion number $N_f$ by (\ref{gs})  
can be estimated treating $\Delta H$ as perturbation.

We write the eigenstates of $H_Q$ and the corresponding energies as
\begin{equation}\label{expa}
|\alpha\rangle=\sum_{i=0}^{\infty}|\alpha{(i)}\rangle,\;
E_{\alpha}=\sum_{i=0}^{\infty} E_\alpha^{(i)},
\end{equation}
where summation over $i$ is with respect to the order of perturbation $\Delta H$.
Then first  order correction to the energy of the lowest-energy state vanishes 
\begin{equation}
E_{gs}^{(1)}=\langle\psi_{gs}^{N_f}{(0)}|\Delta H|\psi_{gs}^{N_f}{(0)}\rangle=0,
\end{equation}
while the first order correction to the state itself is given by 
\begin{equation}\label{corgs}
|\psi_{gs}^{N_f}{(1)}\rangle=\sum_{\beta}|\beta{(0)}
\rangle \langle \beta {(0)}|\frac{\Delta H
}{E_{gs}^{(0)}-E_\beta^{(0)}}|\psi_{gs}^{N_f}{(0)}\rangle,
\end{equation}
where summation is over all states different from the ground state.
To see that correction (\ref{corgs}) is negligible, we 
estimate it's contribution to the energy in the second order of the expansion (\ref{expa}):
\begin{equation}\label{thir}
E_{gs}^{(2)}=\sum_{\beta}
\frac{\langle\psi_{gs}^{N_f}{(0)}|\Delta H|\beta{(0)}\rangle\langle
\beta(0)|\Delta H
|\psi_{gs}^{N_f}(0)\rangle}{E_{gs}^{(0)}-E_\beta^{(0)}},
\end{equation}
with the summation being over all states different from the ground state.
In the sums (\ref{corgs}) and (\ref{thir}) the only non-vanishing contribution  
is from the states of the form 
\ba
&& |kl\rangle=\tilde{B}^\dagger_{kl}
e^{i\frac{g}{\omega^2}(P_{kk}-P_{ll})}\tilde{f}_k^\dagger
\tilde{f}_l|\psi_{gs}^{N_f}(0)\rangle,\;l=1,...,N_f,\;k=N_f+1,...,N,\nn\\
&& E_{kl}^{(0)}=\omega+g (d_l-d_k)+E_{gs}^{(0)}, 
\ea
and therefore for the particular term we have
\ba
&&\left|\langle kl |\Delta H|\psi_{gs}^{N_f}(0)\rangle \right|=
g\left|\sum_{i\neq j}
\langle \psi_{gs}^{N_f}(0)|
\tilde{B}_{kl} \tilde{f}^\dagger_l \tilde{f}_k
\tilde{M}_{ij}\tilde{f}^\dagger_i \tilde{f}_j 
|\psi_{gs}^{N_f}(0)\rangle \right|
=\nn\\
&&=\frac{g}{\sqrt{\omega}}\left|\sum_{i\neq j}
\langle \psi_{gs}^{N_f}(0)|
\delta_{ik}\delta_{jl} \tilde{f}^\dagger_l \tilde{f}_k
\tilde{f}^\dagger_i \tilde{f}_j
|\psi_{gs}^{N_f}(0)\rangle \right|=\frac{g}{\sqrt{\omega}}.
\ea
The second order correction to the lowest-energy state is negligible if condition 
\ba\label{condicija} 
\frac{E_{gs}^{(2)}}{E_{gs}^{(0)}-E_{kl}^{(0)}}\ll1\ea
is fulfilled. Replacing the difference $d_i-d_j$ in $E_{kl}^{(0)}$ in (\ref{condicija})  by order of the average
distance $\sqrt{N/\tilde{\omega}}$ between the lowest $N_f$ and the highest $N-N_f$ levels in the ensemble 
we obtain the condition  
\ba
\frac{E_{gs}^{(2)}}{E_{gs}^{(0)}-E_{kl}^{(0)}}\approx
N_f(N-N_f)\frac{\frac{g^2}{\omega}}{\left(\omega+g
\sqrt{\frac{N}{\tilde{\omega}}}\right)^2}< N^2\frac{\frac{g^2}{\omega}}{\left(\omega+g
\sqrt{\frac{N}{\tilde{\omega}}}\right)^2}\ll 1,
\ea
which can be rewritten as:    
\ba
\frac{\omega^\frac{3}{2}}{Ng}+
\sqrt{\frac{\omega}{N\tilde{\omega}}}\gg1.
\ea
This condition is fulfilled provided that ${g}/{\omega^\frac{3}{2}}\ll 1/N$
or ${{\omega}/{\tilde{\omega}}}\ll N$ which means that the strong coupling limit from our previous paper 
which corresponds to choice $\tilde{\omega}\gg \omega$ and ${g}/{\omega^\frac{3}{2}}\gg \sqrt{N}$ is out of the scope of present
approximation. However, with model introduced in this paper we are able to adjust parameters in such way that aforementioned expected 
displacement $g/\omega^2$ is comparable with scale of the renormalized semicircle $\sqrt{N/\omega_r}$.
For example, we can choose $\omega=N\tilde{\omega}\alpha$, with $\alpha$ taken to be of order unity, in which case  
size of the renormalized semicircle  is approximately $\sqrt{{N}/{\tilde{\omega}}}$. 
Then the condition that displacement $g/\omega^2$ is comparable to the size of the renormalized
semicircle leads to  $g\sqrt{\tilde{\omega}}/(\sqrt{N}\omega^2)=g/(N\sqrt{\alpha}\omega^{3/2}) \sim 1$.

In the next section assuming that parameters are adjusted as described above, 
we use the ground-state wave function ({\ref{gs}) to evaluate the expectation value of 
the level density (\ref{dennf}).

\section{Homolumo gap}
\label{largeg}

We are interested in the expectation value of the density of the $N_f$ lowest levels
in the  state ({\ref{gs}), which is given by\footnote{The subscript $g$ on the left hand
size indicates non-zero fermion-boson interaction strength, compare with (\ref{final}).}:
\ba\label{intint}
&&\lav\rho_{N_f}^<(x)\rav_g^{gs}=
{\cal{N}}(\tilde{\omega},N)\int dM_0 e^{-\tilde{\omega}\mathrm{Tr} M_0^2}
\langle \psi_{gs}^{N_f}(0)| 
\rho_{N_f}^<(x) 
|\psi_{gs}^{N_f}(0)\rangle.
\ea
Due to  dependence of the ground state on the ordering of eigenvalues 
of the matrix $M_0$ we insert the following identity under the integral (\ref{intint}):
\begin{equation}
\sum_{\cal{C}}\prod_{i,\alpha}
\theta(d_{{\cal{C}}(\alpha)}-d_{{\cal{C}}(i)})=1,
\end{equation}
where summation over $\cal{C}$ represents sum over all possible combinations of 
$N_f$ levels from the set of $N$ levels.
Explicitly written we would like to calculate:
\ba\label{intint2}
\lav\rho_{N_f}^<(x)\rav_g^{gs}
&=&
{\cal{N}}(\tilde{\omega},N)
{\cal{N}}(\omega,N)
\sum_{\cal{C}}
\int dM_0 dM 
e^{-\tilde{\omega}\mathrm{Tr} M_0^2}
 e^{-\omega \mathrm{Tr} \left(M-M_0+\frac{g}{2\omega^2}V_c \eta V^\dagger_c
 \right)^2}\times\nn\\ &\times & 
 \prod_{i,\alpha} \theta(d_{{\cal{C}}(\alpha)}-d_{{\cal{C}}(i)})\rho_{N_f}^<(x), 
 \ea
 where unitary matrix $V_c$ maps matrix $M_0$ to the diagonal matrix $D$ with ordered 
 eigenvalues:
 \ba 
 D=V^{\dagger}_cM_0V_c. 
 \ea
 For that purpose we use the following identity:
 \ba\label{heh} 
 \left(\frac{\omega}{\pi}\right)^{N^2/2} e^{-\omega \mathrm{Tr}(M-M_0)^2}&=&
 \sum_k \frac{1}{k!\omega^k} \left.
 \frac{\partial^k}{\partial \epsilon^k} {\cal{N}}(1/\epsilon,N_f)
 e^{-\frac{1}{\epsilon} 
 \mathrm{Tr} (M-M_0)^2}
 \right|_{\epsilon \to 0}=\nonumber\\ &=&
 \sum_k \frac{1}{k! \omega^k} \Delta^k_{M_0} \boldsymbol{\delta}(M-M_0),
 \ea
 where $\boldsymbol{\delta}(M-M_0)$ is defined by 
 \begin{equation}\label{mdf}
 {\boldsymbol{\delta}}(M)=
 \lim_{\omega \to \infty}
 {\cal{N}}(\omega,N)
 e^{-\omega {\rm Tr}M^2}=
 \Pi_{i\leq j} \delta(S_{ij})
 \Pi_{k<l} 
 \delta(A_{kl}),\;S=(M+M^t)/2,\;A=(M-M^t)/2i,\nn
 \end{equation}
while matrix Laplacian $\Delta_{M_0}$ is given by 
\ba
\Delta_{M_0}=\frac{1}{4}\sum_{ij}\frac{\partial^2}{\partial
{M_{0}}_{ij}\partial {M_{0}}_{ji}}. 
\ea 
Then using the identity (\ref{heh}) to expand the second Gaussian in (\ref{intint2}) into powers of $1/\omega$ 
 and after integration by parts we obtain:
\ba\label{deng2}
\lav\rho_{N_f}^<(x)\rav_g^{gs}&=& 
{\cal{N}}(\tilde{\omega},N)\sum_{\cal{C}}
\sum_k \frac{1}{k! \omega^k}\int dM_0 dM
\boldsymbol{\delta}\left(M-M_0+\frac{g}{2\omega^2}V_C\eta V^\dagger_C
\right)\times\nonumber\\
&\times&
\prod_{i,\alpha} \theta(d_{{\cal{C}}(\alpha)}-d_{{\cal{C}}(i)})
\rho_{N_f}^<(x)\Delta^k_{M_0}
e^{-\tilde{\omega}\mathrm{Tr} M_0^2}=\\
&=&
{\cal{N}}({\omega}_r,N)\sum_{\cal{C}}\sum_k
\int dM_0 e^{-\frac{\omega\tilde{\omega}}{\omega+\tilde{\omega}}\mathrm{Tr} M_0^2}
\prod_{i,\alpha} \theta(d_{C(\alpha)}-d_{C(i)})
\delta(x+{g}/{2\omega^2}-d_{C(k)}),\nn
\ea
where in the last line we performed resummation using:  
\ba
\left(\frac{\omega_r}{\pi}\right)^{N^2/2}e^{-\omega_r\mathrm{Tr} M_0^2}=
\left(\frac{\tilde\omega}{\pi}\right)^{N^2/2}\sum_k \frac{1}{k! \omega^k}\Delta^k_{M_0}e^{-\tilde{\omega}\mathrm{Tr} M_0^2}.
\ea
Finally, we can write 
\begin{equation}\label{final1}
\lav\rho_{N_f}^<(x)\rav_g^{gs}=
e^{\frac{g}{2\omega^2}\partial_x}
\lav\rho_{N_F}^<(x)\rav_0^{gs}=\frac{1}{2}{\rm erfc}(\sqrt{\omega_F}
(x+g/2\omega^2 -\mu_F)) \lav\rho(x+{g}/{2\omega^2})\rav_{0}^{gs}.
\end{equation}
Density of empty levels is similarly obtained as 
\ba 
\lav\rho^>_{N-N_f}(x)\rav_{g}^{gs} &=&\frac{1}{2}\left(1+{\rm
erf}(\sqrt{\omega_F}
(x-g/2\omega^2 -\mu_F))\right) \lav\rho(x-{g}/{2\omega^2})\rav_{0}^{gs}.
\ea
In the  expression (\ref{final1}) we notice effect of displacement of the distribution, i.e. the
opening of the gap, and smoothing-out of the edges of the distribution due
to the (complementary) error function.

\section{Discussion and outlook} 
\label{conclusion}

In the previous works \cite{hl1,hl2} we introduced the model having the quantum
mechanical dynamical matrix 
with the minimum in potential energy occurring for the matrix elements being
zero. After the fermion interaction term 
was switched on  we   obtained a reliable homolumo gap  only in the limit of
the large coupling  $g$ of the fermions to the matrix. The resulting spectrum was the one of
two largely separated Wigner semicircles. For the
calculation to be trustworthy and doable we had  
fermion states  either filled or empty but not in superposition. That meant that
we needed the quantum 
fluctuations going as $\sim 1/\sqrt{\omega}$ to be small compared to
the distance in the level energy 
$\sim g/\omega^2$, originating from the push by fermions. 
This could only be achieved  by pulling the filled levels far away from the
empty ones. 

In the present work we, however, introduced yet another mechanism for spreading
the levels  before switching the fermion
interaction on, namely the quenched (or fundamental) randomness of the levels.
The matrix 
elements were by their dynamical potential attracted to a value given 
by  a Gaussian distribution of width   $\tilde{\omega}$. Looking at the case
of $\tilde{\omega}$ being small so that
quenched fluctuations became dominant  we could avoid   the
push becoming larger than the size of the original 
distribution extension, because  the distribution in eigenvalue space was made
extensive by the quenched fluctuations.
Nevertheless we could  still keep the push large compared to the quantum
fluctuations,  so
that our calculations are reliable. Thus we found in
our type of matrix 
model with both quantum and quenched level fluctuation contributions - but in
the case wherein the quenched 
fluctuations dominate - a homolumo-gap effect that meant that we 
\begin{itemize}
\item Start from the no-fermion-interaction spectrum density $\rho$.
\item Fill the lowest levels with $N_f$ fermions. 
\item Push respectively the filled sector and the empty sector
down and up in level-energy by 
${g}/{2\omega^2}$. This operation leaves (in the first approximation) an
interval of length ${g}/{\omega^2}$ in between the 
filled and empty levels in which there are no levels at all. This is the
homolumo gap.
\item In the next approximation we then show that the boundaries of this
level-empty interval are not completely 
sharp but rather smoothed out. The density of levels is  rather  given as the
displaced level density  described in previous point but multiplied by a (complementary) error
function smoothing out the steep level density falls.
Since an error function  is very close to a theta function  in our case, we
effectively have very low level 
density in the gap as long as our approximations is justified.
\end{itemize}
Notice that in the paper we have set $\hbar=1$. We can reintroduce $\hbar$  in our model just using 
\ba
P_{lk}=-i\hbar \frac{\partial}{\partial M_{lk}}.
\ea
We are then free to rescale the energy/Hamiltonian  by $\hbar^2$ and  replace $g\to g/\hbar^2$ and $\omega\to \omega/\hbar$.
From this it is seen that  in the notation of the present paper the classical limit is 
achieved by $\omega \to \infty,\;g\to\infty$, but keeping  $g/\omega^2$ of order one.
This means that the homolumo gap shift, which is just $g/\omega^2$ is actually (possible to be considered) classical effect.
If fact our calculation is based on this limit and thus  (mainly) classical.
Of course letting $\omega\to \infty$ makes the spread of levels be entirely dominated by the quenched
 random distribution and all the levels would classically coincide if it were not for the quenched random $M_0$.

It should be understood that we consider the importance of calculations of the
homolumo-gap effect as being due to the fact that  
the  model assumed  occurs very generally. It is just a model describing some bosonic degrees of
freedom interacting with fermions. We assume a specific 
statistical distribution and the model dynamics, but   we have in mind that
realistic cases of interactions of 
fermions with a system of boson degrees of freedom are usually  too complicated
for analytic analyses.  Therefore it is 
of great significance, if one could obtain some information about such
complicated systems - such as e.g. the homolumo-gap effect - even if one has
modeled 
the complicated system by a random one. The standard example where such a
procedure has been used and is relevant is the case of complex nuclei and
molecules,
as discussed in Introduction.

Furthermore, one could also look at several well-known phenomena
as being in reality homolumo-gap effects although not really always 
announced like that. Mott localization leading to the  transition from
metals to insulators \cite{mott} is the obvious example. 
Using the Hubbard model one may describe a lattice of ``atomic sites'' 
where electrons are sitting  as a model  for some solid crystal. In the Hubbard 
model there is in addition to an effective kinetic term an interaction term 
- with a product of four electron creation or annihilation 
operators - representing a repulsion between electrons sitting on the 
same atom, a Coulomb force representative. 
The Mott insulation transition from a metal occurs for appropriate parameters 
as a consequence of this interaction between the electrons. 
In the Hartree-Fock approximation  to this Hubbard model,
one could imagine that a field ansatz with bosonic degrees of freedom could 
represent the background potentials that were created for the single electrons from
the ``background'' of other electrons. This background field ansatz could now be 
thought of as a system of bosonic degrees of freedom  that could finally be 
represented approximately by our quenched random and dynamical matrix 
$M$. Based on our results,  we could   expect   that a homolumo-gap effect would set in and work towards pushing the 
single electron energy eigenstates away from the neighborhood of the Fermi-surface.
If they are pushed away so that a genuine gap would appear the material would 
go insulating. It is important for such a speculated 
understanding of the mechanism of Mott-insulation to be indeed a 
homolumo-gap effect that there is   a   back reaction  of the electrons 
on the   background considered as bosonic degrees of freedom - in the 
Mott-insulation case the system of electrons itself. 

In principle, the Hartree-Fock approximation can be used to replace direct interaction
between fermions with the effective interaction with background   
described by  the dynamical matrix $M$. Therefore, our general model
could have rather wide range of applications.  
In such a thinking we might even see superconductivity as an example of an
effect of homolumo gap, namely the gap in the quasi-electron spectrum in the 
superconductor state is the homolumo gap.

As a further  development of the present model we plan to  introduce 
a system of conserved or at least approximately conserved quantum numbers 
  for the fermions. One could think of such conserved quantum
numbers  to be   the quasi-momentum components for the electrons in 
a crystal. We could then set up some distribution assumption and some dynamics for
a whole set of matrices $M(\vec{p})$, one for each quasi-momentum vector.
These matrices $M(\vec{p})$  should now be assumed random and dynamical in such
a way that they preserve continuity w.r.t. to the momentum vector $\vec{p}$.
 Neglecting in
first approximation the interaction between the neighboring dynamical matrices
one would  conclude that the spectrum of $M(\vec{p})$
would show a homolumo- gap for every value of the quasi-momentum $\vec{p}$.
In general, however, an attempt to impose the  continuity condition in higher dimensional model   would be obstructed due to topological reasons. 

Let us first think of the restriction of the quantum average of the dynamical 
matrix $M(\vec{p})$ to a surface in the $(n+1)$-dimensional $\vec{p}$ space of topology as an 
sphere $S^n$. Supposing that there were indeed a homolumo gap all along such
a surface the restriction of the quantum-averaged $M(\vec{p})$ along this surface
would be a continuous map of a   $S^n$ sphere into the set of
$N\times N$ matrices having a gap between the lowest $N_f$ eigenvalues and the 
remaining $N-N_f$ ones. The crucial point is that
the space of matrices with this prescribed gap is not
topologically trivial/contractible, but rather can have non-trivial homotopy
group(s). Now the restriction of $M(\vec{p})$ mentioned will represent   an element
in the homotopy group $\Pi_n(\hbox{matrices with the gap})$,
and it   could turn out to be a non-trivial element in this
homotopy group. In that case, however, the quantum-averaged $M(\vec{p})$ could
not be extended to the whole $\vec{p}$-space keeping all over both 
the homolumo gap and the continuity. As a  consequence, we expect  that   there will be some surfaces in the 
quasi-momentum space of $\vec{p}$'s on which the homolumo gap  will no longer remain present after 
imposing continuity. It would 
be near such surfaces with suspended homolumo gap that the low energy excitations
of the (fermion) system would be possible. We hope to refine in the forthcoming article the old 
result \cite{old} for these excitations showing emergent Lorentz invariance. 

Another direction in which we hope to develop the present work is to make the
analogue of the homolumo gap with the fermions replaced by bosons.
Interestingly enough, the model describing interaction of bosons with a
dynamical matrix was discussed as a toy model capable of shading some light 
on the black-hole information problem \cite{Polch}.        
   
\section*{Acknowledgments}
One of us (HBN) would like to thank the Niels Bohr Institute for being allowed
status as emeritus and the Rudjer Bo\v skovi\'c Institute for hospitality there
where most of this work was done.


\begin{thebibliography}{}
\bibitem{jt}
H.~A.~Jahn and E.~Teller, ``Stability of Polyatomic Molecules in Degenerate Electronic States. I. Orbital Degeneracy," Proc.\  Roy.\ Soc.\ London {\bf A161} (1937) 220; \\
M.~Pope and C.~E.~Swenberg, Electronic Processes in Organic Crystals and
Polymers, (2nd ed., Oxford University Press, NY (1999)). 

\bibitem{jt2}
 M.~C.~M.~O'Brien and C.~C.~Chancey, ``The Jahn–Teller effect: An introduction and current review," Am.\ J. \  Phys. {\bf 61} (1993) 688.

\bibitem{Kim}
B.-G. Kim, et al.,
``Energy Level Modulation of HOMO, LUMO, and Band-Gap in Conjugated Polymers for
Organic Photovoltaic Applications", Adv. Funct. Mater.  {\bf 23} (2013) 439.

\bibitem{hl1}
I.~Andri\'c, L.~Jonke, D.~Jurman and H.~B.~Nielsen,
  ``Homolumo Gap and Matrix Model,''
  Phys.\ Rev.\   {\bf D77} (2008) 127701
  [arXiv:0712.3760 [hep-th]].

\bibitem{hl2}
I.~Andri\'c, L.~Jonke, D.~Jurman and H.~B.~Nielsen,
  ``Homolumo Gap from Dynamical Energy Levels,''
  Phys.\ Rev.\  {\bf D80} (2009) 107701
  [arXiv:0909.2346 [hep-th]].
  
\bibitem{Zinn-Justin}
P.~Zinn-Justin, 
"Universality of correlation functions of Hermitian random matrices in an
external field",
Commun. Math. Phys. {\bf 194}  (1998) 631 [arXiv:cond-mat/9705044].

\bibitem{Bleher}
P.~M.~Bleher and A.~B.~J.~Kuijlaars,
"Large n Limit of Gaussian Random Matrices with External Source, Part I",
Commun. Math. Phys.  {\bf 252} (2004) 43 [arXiv:math-ph/0402042];\\
A.~I.~Aptekarev,  P.~M.~Bleher and A.~B.~J.~Kuijlaars,
"Large n Limit of Gaussian Random Matrices with External Source, Part II",
Commun. Math. Phys  {\bf 259} (2005)  367 [arXiv:math-ph/0408041]; \\
P.~M.~Bleher and A.~B.~J.~Kuijlaars,
"Large n Limit of Gaussian Random Matrices with External Source, Part III.
Double Scaling Limit",
Commun. Math. Phys  {\bf 270} (2007) 481 [arXiv:math-ph/0602064].

\bibitem{mehta}
M.~L.~Mehta, "Random Matrices", (2nd Edition, Academic Press, NY(1991)).


\bibitem{gustavson}
J.~Gustavsson, ''Gaussian fluctuations of eigenvalues in the GUE,'' arXiv:math/0401076
[math.PR].

\bibitem{DKMV}
P.~Deift, T.~Kriecherbauer, K.~T.-R McLaughlin, S.~ Venakides and X.~ Zhou,
''Strong asymptotics of orthogonal polynomials with respect to exponential weights,''
Comm. Pure and Appl. Math. {\bf 52} (1999) 1491

  \bibitem{bz}
E.~Br\'ezin and A.~Zee, 
''Universality of the correlations between eigenvalues of large random
matrices,''
Nucl. Phys.  {\bf B402} (1993) 613.

\bibitem{mott}
 N.~F.~Mott, ``Metal-Insulator Transition'', Rev. Mod. Phys.  {\bf 40} (1968)  677.


\bibitem{old}
  C.~D.~Froggatt and H.~B.~Nielsen,
  ``Derivation of Lorentz invariance and three space dimensions in generic field theory,''
  hep-ph/0211106.
 
\bibitem{Polch}
  N.~Iizuka and J.~Polchinski,
  ``A Matrix Model for Black Hole Thermalization,''
  JHEP {\bf 0810} (2008) 028
  [arXiv:0801.3657 [hep-th]].
 
\end{thebibliography}
\end{document}